\def\arxiv{1}    

\let\accentvec\vec 
\documentclass{llncs}
\let\vec\accentvec %

\ifnum\arxiv=1
\fi

\usepackage[utf8]{inputenc}
\usepackage[T1]{fontenc}
\usepackage{microtype}

\usepackage{amssymb}
\usepackage{amsmath}
\usepackage{amsfonts}

\newcommand{\proofEnd}{{\hfill\ensuremath{\blacksquare}}}
\newenvironment{proof_claim}{\upshape{\textsc{Proof of Claim.}}} {\hfill\ensuremath{\square}\smallskip}
\spnewtheorem{myClaim}{Claim}{\itshape}{\upshape}

\newcommand{\figPath}{.}
\usepackage{url}
\usepackage{hyperref}
\usepackage{xcolor}
\definecolor{darkblue}{rgb}{0,0,.5}
\hypersetup{colorlinks=true, breaklinks=true, linkcolor=darkblue, menucolor=darkblue, urlcolor=darkblue, citecolor=darkblue}

\usepackage{graphicx}
\usepackage{grffile}
\usepackage{subfigure}
\usepackage{sidecap}

\usepackage[neveradjust]{paralist}

\newcommand{\degr}{\ensuremath{d}}
\newcommand{\randv}{\ensuremath{D}}
\newcommand{\keys}{\ensuremath{n}}
\newcommand{\cells}{\ensuremath{m}}
\newcommand{\pmf}{\ensuremath{{\rho}}}
\newcommand{\mean}{{\mathrm{\scriptstyle\Delta}}}
\newcommand{\Amean}{\bar{\mean}}
\newcommand{\E}{\mathrm{E}}

\newcommand{\floor}[1]{\ensuremath{\lfloor #1\rfloor}}
\newcommand{\ceil}[1]{\ensuremath{\lceil #1\rceil}}
\newcommand{\eps}{\ensuremath{\varepsilon}}
\newcommand{\blank}{\text{ }}
\newcommand{\graph}{\ensuremath{G}}
\newcommand{\graphWR}{\ensuremath{\tilde{G}}}
\newcommand{\low}{\ensuremath{l}}
\newcommand{\high}{\ensuremath{k}}
\newcommand{\abs}[1]{\lvert#1\rvert}

\newcommand{\bfrac}[2]{ \left(\frac{#1}{#2}\right) }

\newcommand{\evNM}{\bar{\mathcal{M}}}
\newcommand{\evM}{{\mathcal{M}}}

\newcommand{\evP}{{\mathcal{BI}}}
\newcommand{\evA}{{\mathcal{A}}}
\newcommand{\evB}{{\mathcal{B}}}
\newcommand{\Fail}{\mathrm{Fail}}
\newcommand{\fail}{\mathrm{fail}}
\newcommand{\bl}{\beta}
\newcommand{\hf}{\gamma}
\newcommand{\mgap}{\hspace{-1em}}

\begin{document}
 \abovedisplayskip0.12cm plus0.12cm minus0.12cm
 \belowdisplayskip0.12cm plus0.12cm minus0.12cm

\title{Towards Optimal Degree-distributions for Left-perfect Matchings in\texorpdfstring{\\}{}Random Bipartite Graphs}
\author{%
Martin Dietzfelbinger\thanks{Research supported by DFG grant DI 412/10-2.} \and Michael Rink$^{\star}$}

\institute{%
Fakultät für Informatik und Automatisierung, Technische Universität Ilmenau
\email{\{martin.dietzfelbinger,michael.rink\}@tu-ilmenau.de}
}

\maketitle
\begin{abstract}
\vspace{-0.4cm}
Consider a random bipartite multigraph $\graph$ with $\keys$ left nodes and $\cells \geq \keys\geq2$
right nodes. Each left node $x$ has $\degr_x\geq1$ random right neighbors.
The average left degree $\Amean$ is fixed, $\Amean\geq2$.
We ask whether for the probability that $\graph$ has a left-perfect matching
it is advantageous not to fix $\degr_x$ for each left node $x$ but rather choose it
at random according to some (cleverly chosen) distribution. We show the following,
provided that the degrees of the left nodes are independent: 
If $\Amean$ is an integer then it is optimal to use a fixed degree of $\Amean$ for all left nodes.
If $\Amean$ is non-integral then an optimal degree-distribution has the property that
each left node $x$ has two possible degrees, $\floor{\Amean}$ and $\ceil{\Amean}$,
with probability $p_x$ and $1-p_x$, respectively, where $p_x$ is from the closed interval~$[0,1]$
and the average over all $p_x$ equals $\ceil{\Amean}-\Amean$.
Furthermore, if $\keys=c\cdot \cells$ and $\Amean>2$ is constant,
then each distribution of the left degrees that meets the conditions above determines the same
threshold $c^*(\Amean)$ that has the following property as $\keys$ goes to infinity:
If $c<c^*(\Amean)$ then there exists a left-perfect matching with high probability.
If $c>c^*(\Amean)$ then there exists no left-perfect matching with high probability.
The threshold $c^*(\Amean)$ is the same as the known threshold for offline $k$-ary cuckoo hashing
for integral or non-integral $k=\Amean$.
\vspace{-0.1cm}
\end{abstract}

\section{Introduction}
We study bipartite multigraphs $\graph$ with left node set $S$ and right node set ${T}$,
where each left node $x$ from $S$ has $\randv_x$ right neighbors.
The right neighbors are chosen at random with replacement from ${T}$,
where the number of choices $\randv_x$ is a random variable that follows some probability mass function $\pmf_x$.
Let $\abs{S}=n$ and let $\abs{{T}}=m$ as well as $1\leq\randv_x\leq \cells$ for all $x$ from $S$.
For each $x$ from $S$ let $\mean_x$ be the mean of $\randv_x$, that is, $\mean_x=\sum_{\low=1}^\cells \low\cdot \rho_x(\low)$,
and let $\Amean$ be the average mean, i.e., $\Amean={1}/{n}\cdot \sum_{x\in S} \mean_x$.
We assume that the random variables $\randv_x, x\in S,$ are independent and $\Amean$ is a given constant. 

Our aim is to determine a sequence of probability mass functions $(\pmf_x)_{x\in S}$ for the random variables $(D_x)_{x\in S}$
that maximizes the probability that the random graph $\graph=\graph\big(\Amean, (\pmf_x)_{x\in S} \big)$
has a matching that covers all left nodes, i.e., a left-perfect matching\footnote{In the following we will use ``matching'' and ``left-perfect matching'' synonymously.}.
We call such a sequence \emph{optimal}. Note that there must be some optimal sequence for compactness reasons.


\subsection{Motivation and Related Work}
Studying irregular bipartite graphs has lead to major improvements 
in the performance of erasure correcting codes.
For example in \cite{LMSS_Tornado_2001} Luby et al. showed how to increase the
fraction of message bits that can be recovered 
for a fixed number of check bits by using carefully
chosen degree sequences for both sides of the underlying bipartite graph.
The recovery process for erased message bits 
translates directly into a greedy algorithm 
for finding a matching in the bipartite graph associated with the recovery process.
This was the motivation for the authors of~\cite{DGMMPR_tight_2009_full,DGMMPR_tight_2010}
to study irregularity in the context of offline $k$-ary cuckoo hashing.
Here one has a bipartite graph with left nodes corresponding to
keys and right nodes corresponding to table cells,
where each key randomly chooses table cells without replacement and 
the aim is essentially to find a left-perfect matching.
In \cite{DGMMPR_tight_2009_full} it was proven that if the degree of each left node 
follows some distribution with identical mean
and is independent of the other nodes then 
it is optimal in an asymptotic sense if
the degree of each left node is concentrated around its mean.
This is in contrast of the following observation 
in~\cite{R_mixed_preparation} in analogy to \cite{LMSS_Tornado_2001}:
an uneven distribution of the degrees of the left nodes
can increase the probability for the existence of a matching that has the advantage that it can be calculated
in linear time, by successively assigning left nodes to right nodes of degree one and removing them from the graph.

\subsection{Results}
We will show that for given parameters $\keys,\cells$, and $\Amean$ there is an optimal sequence
of probability mass functions that concentrates the degree of the left nodes
around $\floor{\Amean}$ and $\ceil{\Amean}$.
Furthermore, if $\Amean$ is an integer we can explicitly determine this optimal sequence. In the
case that $\Amean$ is non-integral we will identify a tight condition that an optimal sequence must meet. 
\begin{theorem}
\label{theo:main}
Let $\keys\leq\cells$, as well as $\keys,\Amean \geq 2$, and
let $({\pmf}_x)_{x\in S}$ be an optimal sequence for parameters $(\keys,\cells,\Amean)$.
Then the following holds for all $x \in S$.
\begin{compactenum}[(i)]
\item If $\Amean$ is an integer, then ${\pmf}_x(\Amean)=1$.
\item If $\Amean$ is non-integral, 
then $\pmf_x(\floor{\Amean})\in[0,1]$ and ${\pmf}_x(\ceil{\Amean})=1-{\pmf}_x(\floor{\Amean})$.
\end{compactenum}
\end{theorem}
The second statement is not entirely satisfying since it identifies no optimal solution.
However, we will give strong evidence that in the situation of Theorem~\ref{theo:main}~$(ii)$ there is no single,
simple description of a distribution that is optimal for all feasible node set sizes.

Since the case $\Amean=2$ is completely settled by Theorem~\ref{theo:main}~$(i)$,
we focus on the cases where $\Amean>2$, with the additional condition that the number of left nodes is linear in the number of right nodes, that is $n=c\cdot m$ for constant $c>0$.
We show that for sufficiently large $n$ all sequences that meet the condition of Theorem~\ref{theo:main}~$(ii)$ 
asymptotically lead to the same matching probability. Therefore, we call these sequences \emph{near optimal}.
\begin{proposition}
\label{prop:thresholds}
Let $n=c\cdot m$, for constant $c>0$, and let $(\pmf_x)_{x\in S}$ be a near optimal sequence with average expected degree $\Amean>2$.
Then for sufficiently large $n$ there is a threshold $c^*(\Amean)$ such that the random graph $\graph=\graph\big(\Amean, (\pmf_x)_{x\in S} \big)$ 
has the following property. 
\begin{compactenum}[(i)]
\item  If  $c<c^*(\Amean)$, then  $\graph$ has a matching with probability $1-o(1)$.
\item  If  $c>c^*(\Amean)$, then   $\graph$ has no matching with probability $1-o(1)$.
\end{compactenum}
\end{proposition}
The threshold $c^*(\Amean)$ is exactly the same as the threshold given in the context of $k$-ary cuckoo hashing for
integral $k$ \cite{FM_maximum_2009,FP_orientability_2010,DGMMPR_tight_2010}, and non-integral $k$ \cite{DGMMPR_tight_2010}, where $k=\Amean$.

So in the case that $n=c\cdot m$ all near optimal sequences are hardly distinguishable in terms
of matching probability, at least asymptotically, but we will give strong evidence that there are only two sequences that
can be optimal, where the decision which one is the optimal one depends on the ratio $c$.
\begin{conjecture}
\label{con:optimal}
Let $({\pmf}_x)_{x\in S}$ be an \emph{optimal sequence} for parameters $(\keys,\cells,\Amean)$ 
in the situation of Theorem~\ref{theo:main}~$(ii)$ for $n=c\cdot m$ and constant $c>0$ and 
$\Amean>2$. Let $\alpha=\ceil{\Amean}-\Amean$.
\begin{compactenum}[$(i)$]
\item If  $c<c^*(\Amean)$, then $\pmf_x(\floor{\Amean})=1$ for $\alpha\cdot n$ nodes and $\pmf_x(\ceil{\Amean})=1$ for $(1-\alpha)\cdot n$ nodes 
(assuming that $\alpha\cdot n$ is an integer).
\item If  $c>c^*(\Amean)$, then $\pmf_x(\floor{\Amean})=\alpha$ and $\pmf_x(\ceil{\Amean})=1-\alpha$  for all $x\in S$.
\end{compactenum}
\end{conjecture}
That is, if $c$ is to the left of the threshold then it is optimal to fix the degrees of the left nodes, and
if $c$ is to the right of the threshold then it is optimal to let each left node choose its degree at random
from $\floor{\Amean}$ and $\ceil{\Amean}$, by identical, independent experiments.

\smallskip
\noindent\textbf{Overview of the paper}
The next section, which is also the main part, covers the proof of Theorem~\ref{theo:main}.
It is followed by a section devoted to the discussion of Conjecture~\ref{con:optimal}.
The proof of Proposition~\ref{prop:thresholds} is given in 
\ifnum\arxiv=1
Appendix~\ref{app:proposition}, since it is only using 
\else
the full version of this paper~\cite[Appendix B]{full_version}. It uses 
\fi
standard techniques on concentration bounds for nodes of certain degrees. 

\section{Optimality of Concentration in a Unit Length Interval}
In this section we prove Theorem~\ref{theo:main}.
We define the \emph{success probability} of a random graph as the probability that this graph has a matching.
Let $n,m$ and $\Amean$ be fixed and consider some arbitrary but fixed sequence of probability mass functions $(\pmf_x)_{x\in S}$.
We will show that if this sequence has certain properties then we can do a modification, obtaining a new sequence $(\pmf'_x)_{x\in S}$ 
with the same average expected value $\Amean$, such that $\graph\big( \Amean, (\pmf'_x)_{x\in S}\big)$ has 
a strictly higher success probability than $\graph\big( \Amean, (\pmf_x)_{x\in S}\big)$.
\begin{lemma}[{Variant of \cite[Proposition 4]{DGMMPR_tight_2010}}]
\label{lem:concentration_around_mean}
Let $(\pmf_x)_{x\in S}$ be given.
Let $z\in S$ be arbitrary but fixed.
If in $\pmf_z$ two degrees with distance at least $2$ have nonzero probability then $(\pmf_x)_{x\in S}$ is not optimal.
\end{lemma}
The lemma was stated in \cite{DGMMPR_tight_2010} and proven in \cite{DGMMPR_tight_2009_full} for a slightly different graph model.
Its proof runs along the lines of \cite{DGMMPR_tight_2009_full}; it is 
\ifnum\arxiv=1
included in Appendix~\ref{app:lemma_concentration} for the convenience of the reader.
\else
given in~\cite[Appendix A]{full_version}.
\fi
After applying the first lemma repeatedly one sees that in an optimal sequence
each left node node has either a fixed degree (with probability 1) or two possible degrees with non-zero probability,
where these degrees differ by~1.
The lemma and~\cite{DGMMPR_tight_2009_full,DGMMPR_tight_2010} do not say anything about the relation between the degrees of different nodes.
This follows next.
\begin{lemma}
\label{lem:distance_at_most_2}
Let $(\pmf_x)_{x\in S}$ be given, where for each $x\in S$ the only degrees with 
nonzero probability are from $\{\floor{\mean_x},\ceil{\mean_x}\}$.
Let $y,z\in S$ be arbitrary but fixed.
If $\floor{\mean_y}$ and $\floor{\mean_z}$ have distance at least~$2$, 
or $\ceil{\mean_y}$ and $\ceil{\mean_z}$ have distance at least~$2$,
then $(\pmf_x)_{x\in S}$ is not optimal.
\end{lemma}
Lemma~\ref{lem:distance_at_most_2} is proved in Section~\ref{sec:lemma_distance_at_most_2}.
Using Lemma~\ref{lem:distance_at_most_2} one concludes that an optimal sequence restricts the means $\mean_x$, for each $x\in S$,
to an open interval $(\low-1,\low+1)$ for some integer constant $\low\geq 2$. 
Hence all degrees that appear with non-zero probability must be from $\{\low-1,\low,\low+1\}$.
With the help of the next lemma one concludes that actually two values are enough.
\begin{lemma}
\label{lem:distance_2}
Let $(\pmf_x)_{x\in S}$ be given, where for each $x\in S$ the only degrees with 
nonzero probability are from $\{\floor{\mean_x},\ceil{\mean_x}\}$. 
Let $y,z\in S$ be arbitrary but fixed and assume that $\mean_y$ and $\mean_z$ are non-integral.
If $\ceil{\mean_y}$ and $\floor{\mean_z}$ have distance $2$ then $(\pmf_x)_{x\in S}$ is not optimal.
\end{lemma}
Lemma~\ref{lem:distance_2} is proved in Section~\ref{sec:lemma_distance_2}.
Combining Lemmas~\ref{lem:concentration_around_mean},~\ref{lem:distance_at_most_2}, and \ref{lem:distance_2},
we obtain the following for an optimal sequence.
If $\low \leq \Amean <\low +1$ then it holds $\low \leq \mean_x \leq \low+1$, for all $x \in S$,
and all degrees that appear with non-zero probability must be from $\{\low,\low+1\}$.
If $\Amean$ is an integer, then by definition of $\Amean$, we have $\pmf_x(\Amean)=1$ for all $x\in S$.
Hence Theorem~\ref{theo:main} follows.

So, to complete the proof of the theorem, it remains to show the three lemmas, which
is done in the following two sections for Lemmas~\ref{lem:distance_at_most_2} and \ref{lem:distance_2}, and 
in
\ifnum\arxiv=1
Appendix~\ref{app:lemma_concentration}
\else
\cite[Appendix A]{full_version}
\fi
for Lemma~\ref{lem:concentration_around_mean}.
We make use of the following definitions.

For each set $S'\subseteq S$ let $\graph_{S'}$ be the induced bipartite subgraph of $\graph$ with
left node set $S'$ and right node set ${T}$, particularly $\graph_S=\graph$.
A matching in $\graph_{S'}$ is a matching that covers all left nodes (left-perfect matching).
We define $\evM_{S'}$ as the event that $\graph_{S'}$ has a matching.
\subsection{Average Degrees of Different Nodes are Close}
\label{sec:lemma_distance_at_most_2}
In this section we prove Lemma~\ref{lem:distance_at_most_2}.
Consider the probability mass functions $\pmf_y$ and $\pmf_z$ for the degrees $\randv_y$ and $\randv_z$ respectively.
By the hypothesis of the lemma, $\pmf_y$ and $\pmf_z$ are concentrated on two values each, i.e.,
\begin{align*}
\pmf_y(\high)=p, \ \pmf_y(\high+1)=1-p & & \pmf_z(\low) =q, \ \pmf_z(\low+1) =1-q \blank,
\end{align*}
with $p\in[0,1]$ and $q\in[0,1]$. By the assumption, we may arrange things so that $\high-\low\geq 2$ and
\begin{compactitem}
\item[$(i)$] $\high  =\floor{\mean_y}$, $\low  =\floor{\mean_z}$ as well as $p=1-(\mean_y-\floor{\mean_y})$, $q=1-(\mean_z-\floor{\mean_z})$,
\item[or $(ii)$] $\high+1=\ceil{\mean_y}$,  $\low+1=\ceil{\mean_z}$ as well as $p=\ceil{\mean_y}-\mean_y$, $q=\ceil{\mean_z}-\mean_z$.
\end{compactitem}\smallskip
We will show that changing $\pmf_y$ to $\pmf'_y$
and $\pmf_z$ to $\pmf'_z$ such that $\mean_y'=\mean_y-1$ and $\mean_z'=\mean_z+1$, via
\begin{align*}
\pmf'_y(\high-1) =p, \ \pmf'_y(\high)=1-p && \pmf'_z(\low+1)=q, \ \pmf'_z(\low+2)=1-q \blank,
\end{align*}
will strictly increase the probability that $\graph_S$ has a matching, while it does not change $\Amean$. For this, will show
\begin{align*}
\Pr\left( \evNM_S \mid \pmf_y,\pmf_z \right) > \Pr\left( \evNM_S \mid \pmf'_y, \pmf'_z\right) \blank,
\end{align*}
abusing condition notation a little to indicate changed probability spaces.
We fix the neighborhood $N_x$ for the remaining elements $x \in S-\{y,z\}$ and therefore the graph $\graph_{S-\{y,z\}}$.
Since there can be a matching for $S$ only if there is a matching for $S-\{y,z\}$ it is sufficient to show that
\begin{align}
\label{eq:failure_lemma_2}
\Pr\left( \evNM_S \mid \evM_{S-\{y,z\}}, \pmf_y, \pmf_z \right) > \Pr\left( \evNM_S \mid \evM_{S-\{y,z\}}, \pmf'_y, \pmf'_z\right) \blank.
\end{align}
Let $\Fail(\degr_y,\degr_z)= \Pr\left( \evNM  \mid \evM_{S-\{y,z\}} , \randv_y=\degr_y, \randv_z=\degr_z \right)$. Then \eqref{eq:failure_lemma_2}
holds if and only if
 \begin{align}
  \label{eq:require_lemma_2}
   \sum_{\substack{\degr_y\in\{\high,\high+1\}\\\degr_z\in\{\low,\low+1\}}}\mgap   \Fail(\degr_y,\degr_z) \cdot \rho_y(\degr_y)\cdot\rho_z(\degr_z) 
 >\mgap \sum_{\substack{\degr_y\in\{\high-1,\high\}\\\degr_z\in\{\low+1,\low+2\}}}\mgap \Fail(\degr_y,\degr_z) \cdot \rho'_y(\degr_y)\cdot\rho'_z(\degr_z) \blank. 
 \end{align}
Note that if $\high-\low=2$ then the summand regarding $\degr_y= \high$ and $\degr_z=\low+1$ on the left-hand side
is the same as the summand regarding $\degr_y= \high-1$ and $\degr_z=\low+2$ on the right-hand side.
Hence, to prove \eqref{eq:require_lemma_2} it is sufficient to show that
\begin{align}
\label{eq:failure_lemma_2_simple}
  \Fail(\high,\low)>\Fail(\high-1,\low+1)\blank.
\end{align}
For this, consider the fixed graph $\graph_{S-\{y,z\}}$. We classify the right nodes of $\graph_{S-\{y,z\}}$ according to the following three types:
\begin{compactitem}
 \item We call $v$ \emph{blocked}   if $v$ is matched in all matchings of $\graph_{S-\{y,z\}}$.
 \item We call $v$ \emph{free}      if $v$ is never matched in any matching of $\graph_{S-\{y,z\}}$.
 \item We call $v$ \emph{half-free} if $v$ is neither a blocked nor a free node.
\end{compactitem}
Let $B$ be the set of blocked nodes, let $F$ be the set of free nodes, and let $\mathit{HF}$ be the set of half-free nodes.
Elements of $\bar{B}=F \cup \mathit{HF}$ are called \emph{non-blocked} nodes.
For a moment consider only the non-blocked nodes. For each right node set $V\subseteq \bar{B}$
let $H_V$ be an auxiliary graph with node set $V$ that
has an edge between two nodes $v_1,v_2 \in V$ if and only if there exists a matching
for $\graph_{S-\{y,z\}}$ in which $v_1$ and $v_2$ simultaneously are \emph{not} matched.
Let $V$ be an arbitrary but fixed subset of $\bar{B}$. The following observation is crucial.
\begin{myClaim}
\label{claim:H_connected}
If $H_V$ has any edges at all then it is connected.
\end{myClaim}
\begin{proof_claim}
First note that if there is a free node in $V$ then $H_V$ is connected by definition of the edge set of $H_V$.
Therefore it remains to consider the case where all nodes of $V$ are half-free nodes.
It is sufficient to show that if for three nodes $v_1, v_2,v_3$ from $\mathit{HF}$ the edge $(v_1,v_2)$
is in $H_V$ then one of the edges $(v_1,v_3)$ or $(v_2,v_3)$ must be present as well.
Assume for a contradiction $(v_1,v_2)$ is an edge but $v_3$ is neither adjacent to $v_1$ nor to $v_2$.
This implies that there are two matchings in $\graph_{S-\{y,z\}}$, $M$ and $M'$ say,
such that in $M$
\begin{compactitem}
 \item node $v_3$ is unmatched ($v_3$ is a non-blocked node), but 
 \item nodes $v_1$ and $v_2$ are matched since edges $(v_1,v_3)$ and $(v_2,v_3)$ are not in $H_V$,
\end{compactitem}
and in $M'$ we have:
\begin{compactitem}
 \item node $v_3$ is matched ($v_3$ is a half-free node), but
 \item $v_1$ and $v_2$ are unmatched since edge $(v_1,v_2)$ is in $H_V$.
\end{compactitem}
Now consider the bipartite multigraph $M\cup M'$ consisting of all edges from both matchings and the corresponding nodes. The graph $M\cup M'$ has the following properties:
Nodes on the left side have degree 2 (both matchings are left-perfect). Nodes on the right side have degree 1 or 2, in particular, $v_1$,$v_2$,$v_3$ have degree 1. Hence $M\cup M'$ has only paths and cycles of even length. On all paths and cycles edges from $M$ and $M'$ alternate.
Nodes $v_1$ and $v_2$ must be at the ends of two distinct paths (since both are incident to $M$-edges). Node $v_3$ must be at the end of a path (incident to an $M'$-edge).

Without loss of generality, we may assume that $v_1$ and $v_3$ do not lie on the same path.
Starting from $M'$, we get a new matching in which neither $v_1$ nor $v_3$ are matched by
replacing the $M'$-edges on the path with $v_3$ by the $M$-edges on this path.
Therefore there must be an edge $(v_1,v_3)$ in $H_V$, which contradicts our assumption, proving the claim.
\end{proof_claim}

\noindent Now consider the set $\bar{B}$ of non-blocked nodes and the corresponding graph $H_{\bar{B}}$. 
We define $\sim$ as the following binary relation: $v_1\sim v_2$, for nodes $v_1$ and $v_2$, if
$(v_1,v_2)$ is not an edge in $H_{\bar{B}}$.
\begin{myClaim}
The relation $\sim$ (no edge) is an equivalence relation.
\end{myClaim}
\begin{proof_claim}
Clearly $\sim$ is reflexive and symmetric. Assume for a contradiction $\sim$ is not transitive.
That is, we have three nodes $v_1,v_2$ and $v_3$ with $v_1\sim v_2$ and $v_2\sim v_3$ but $v_1\not \sim v_3$.
Let $V=\{v_1,v_2,v_3\}$. Since $v_1\not \sim v_3$, the edge $(v_1,v_3)$ is in $H_{\bar{B}}$ and therefore in $H_V$.
According to Claim \ref{claim:H_connected} $H_V$ must be connected, i.e., $H_V$ and therefore $H_{\bar{B}}$ must contain $(v_1,v_2)$ or $(v_2,v_3)$.
Hence $v_1\not \sim v_2$ or $v_2\not\sim v_3$, which is a contradiction.
\end{proof_claim}

\noindent According to the claim it follows that the right node set ${T}$ of $\graph_{S-\{y,z\}}$ can be subdivided into
disjoint segments $B \cup I_1 \cup I_2 \cup \ldots ={T}$, where $B$ is the set of blocked nodes and $I_1,I_2,\ldots$ are
the maximal independent sets in $H_{\bar{B}}$
and the equivalence classes of $\sim$, respectively.
For each pair $I_s,I_t$, with $s\neq t$, it holds that $H_{I_s\cup I_t}$ is a complete bipartite graph.
Note that each free node leads to a one-element set $I_s$.
With this characterization of $H_{\bar{B}}$ we can express the event that for 
a fixed neighborhood $N_x$, $x\in S-\{y,z\}$, which admits a matching for
$\graph_{S-\{y,z\}}$, there is no matching for $\graph_S$ as follows
\begin{align}
\label{eq:characterization_lemma_2}
\{ N_y \subseteq B \}  \cup  \{N_z \subseteq B\}   \cup  \bigcup_{j} \{(N_y\cup N_z) \subseteq (B\cup I_j)\}  \blank.
\end{align}
Let $\evP_{S-\{y,z\}}(b,r,i_1,\ldots,i_r)$ be the event that $\graph_{S-\{y,z\}}$
has $\abs{B}=b$ many blocked nodes and $r$ (nonempty) maximal independent sets according to the definition above,
with $\abs{I_j}=i_j$ and $i_1\leq i_2\leq \ldots\leq i_r$.
Let 
\begin{equation*}
\begin{split}
 \fail&({\degr_y},{\degr_z},b,r,i_1,\ldots,i_r)=\\
& \Pr\left( \evNM  \mid \evM_{S-\{y,z\}}, \randv_y={\degr_y}, \randv_z={\degr_z}, \evP_{S-\{y,z\}}(b,r,i_1,\ldots,i_r) \right) \blank.
\end{split}
\end{equation*}
Then \eqref{eq:characterization_lemma_2} implies that
\begin{align*}
\fail({\degr_y},{\degr_z},b,r,&i_1,\ldots,i_r)
=\bfrac{b}{\cells}^{\degr_y}  + \bfrac{b}{\cells}^{\degr_z} - \bfrac{b}{\cells}^{\degr_y} \cdot \bfrac{b}{\cells}^{\degr_z} \\
&+\sum_{j=1}^r \left[ \bfrac{i_j+b}{\cells}^{\degr_y} - \bfrac{b}{\cells}^{\degr_y} \right]\cdot 
             \left[ \bfrac{i_j+b}{\cells}^{\degr_z}  - \bfrac{b}{\cells}^{\degr_z} \right] \blank.
\end{align*}
Using the law of total probability we can rewrite the value $\Fail({\degr_y},{\degr_z})$ (line below \eqref{eq:failure_lemma_2})
as follows:
\begin{equation*}
\begin{split}
\Fail({\degr_y},{\degr_z})=\sum\limits_{(b,r,i_1,\ldots,i_r)}\fail&({\degr_y},{\degr_z},b,r,i_1,\ldots,i_r) \\[-3ex]
                                                               \cdot&  \Pr(\evP_{S-\{y,z\}}(b,r,i_1,\ldots,i_r) \mid \evM_{S-\{y,z\}}) \blank.
\end{split}
\end{equation*}
We will abbreviate $\fail({\degr_y},{\degr_z},b,r,i_1,\ldots,i_r)$ by $\fail({\degr_y},{\degr_z})$
for the rest of the paper. In order to prove \eqref{eq:failure_lemma_2_simple} it is sufficient to show
\begin{align}
\label{eq:fail_lemma_2}
  \fail(\high,\low)>\fail(\high-1,\low+1)\blank,
\end{align}
for each $\evP$-vector $(b,r,i_1,\ldots,i_r)$.
Let $\hf_j=i_j/\cells$ and let $\bl=b/\cells$. Thus,
\begin{align}
\label{eq:fail}
 \fail(\high,\low)=\bl^\high + \bl^\low - \bl^{\high+\low} + \sum_{j=1}^r \left[(\hf_j+\bl)^\high-\bl^\high \right] \cdot\left[(\hf_j+\bl)^\low-\bl^\low \right] \blank.
\end{align}
Hence, inequality \eqref{eq:fail_lemma_2} holds if and only if 
\begin{align}
\label{eq:fail_left_and_right}
 \bl^\high + \bl^\low - \bl^{\high-1} - \bl^{\low+1}  >&\sum_{j=1}^r \left[(\hf_j+\bl)^{\high-1}-\bl^{\high-1} \right] \cdot\left[(\hf_j+\bl)^{\low+1}-\bl^{\low+1} \right] \notag\\
                                   &         - \left[(\hf_j+\bl)^\high-\bl^\high \right] \cdot\left[(\hf_j+\bl)^\low-\bl^\low \right]\notag\\
\Leftrightarrow \hspace{0.5cm}
(1-\bl)\cdot (\bl^\low -\bl^{\high-1})>& \sum_{j=1}^r \hf_j\cdot \underbrace{\left[\bl^\low \cdot(\hf_j+\bl)^{\high-1} -\bl^{\high-1}\cdot(\hf_j+\bl)^\low  \right]}_{\phi(\low,\high,\hf_j,\bl)} \blank.
\end{align}
Note that if $r=1$ there is no matching for $\graph_S$. Hence we are only interested in the case $r\geq2$, which
implies that $i_j<\cells-b$ and $\hf_j<1-\bl$, respectively.
Consider the right-hand side of the inequality. The expression within the square brackets increases monotonically with increasing $\hf_j$,
since we have
\begin{align*}
 \frac{\partial\phi(\low,\high,\hf_j,\bl)}{\partial \hf_j}=&(\high-1)\cdot \bl^\low \cdot(\hf_j+\bl)^{\high-2}-\low\cdot \bl^{\high-1}\cdot(\hf_j+\bl)^{\low-1} \overset{!}{>}0\\
\Leftrightarrow  \ & \frac{\high-1}{\low} \cdot (\hf_j+\bl)^{\high-\low-1} > \bl^{\high-\low-1} \blank,
\end{align*}
and the last inequality holds because of $\high-\low \geq 2$ and $\hf_j+\bl > \bl$.
Therefore replacing $\hf_j$ with $1-\bl$ within $\phi$ and using that $\sum_{j=1}^r \hf_j = 1-\bl$
strictly increases the right-hand side of \eqref{eq:fail_left_and_right} and yields the left-hand side of \eqref{eq:fail_left_and_right}.
But since we assume $\hf_j<1-\bl$ the strict inequality holds.
Due to the fact that the event $\{r\geq 2\}$ has positive probability Lemma~\ref{lem:distance_at_most_2} follows.\proofEnd

\subsection{Optimal Distributions Use Only Two Neighboring Degrees}
\label{sec:lemma_distance_2}
In this section we prove Lemma~\ref{lem:distance_2}.
Consider the probability mass functions $\pmf_y$ and $\pmf_z$ for the degrees $\randv_y$ and $\randv_z$ respectively.
Let $\floor{\mean_y}=\low$ and $\floor{\mean_z}=\low-1$ as well as $p=1-(\mean_y-\floor{\mean_y})$ and $q=1-(\mean_z-\floor{\mean_z})$.
By the hypothesis of the lemma we have 
\begin{align*}
\pmf_y(\low)  =p, \ \pmf_y(\low+1)=1-p && \pmf_z(\low-1)=q, \ \pmf_z(\low)  =1-q \blank,
\end{align*}
with $p\in(0,1)$ and $q\in(0,1)$. 
To prove Lemma~\ref{lem:distance_2} we will show that
changing $\pmf_y$ to $\pmf'_y$ and $\pmf_z$ to $\pmf'_z$, via
\begin{align*}
\pmf'_y(\low)=p+\eps, \ \pmf'_y(\low+1) =1-p-\eps && \pmf'_z(\low-1)=q-\eps, \ \pmf'_z(\low)=1-q+\eps \blank,
\end{align*}
for some small perturbation $\eps \neq 0$ will strictly increase the probability that $\graph_S$ has a matching, while it does not change $\Amean$. 
Hence as in the proof of Lemma~\ref{lem:distance_at_most_2} we will show that
\begin{align*}
\Pr\left( \evNM_S \mid \pmf_y,\pmf_z \right) > \Pr\left( \evNM_S \mid \pmf'_y, \pmf'_z\right) \blank.
\end{align*}
As before we fix the neighborhood $N_x$ for the remaining elements $x \in S-\{y,z\}$ and therefore the graph $\graph_{S-\{y,z\}}$.
As in Lemma~\ref{lem:distance_at_most_2} we conclude that it is sufficient to show that for some perturbation term $\eps \neq 0$ we have
\begin{align*}
   \sum_{\substack{\degr_y\in\{\low,\low+1\}\\\degr_z\in\{\low-1,\low\}}} \Fail(\degr_y,\degr_z) \cdot \rho_y(\degr_y)\cdot\rho_z(\degr_z) 
 &> \sum_{\substack{\degr_y\in\{\low,\low+1\}\\\degr_z\in\{\low-1,\low\}}} \Fail(\degr_y,\degr_z) \cdot \rho'_y(\degr_y)\cdot\rho'_z(\degr_z) \blank. 
\end{align*}
Subtracting the left-hand side from right-hand side gives
\begin{align}
\label{eq:failure_lemma_3_eps}
 &&\left[ -\eps^2-\eps\cdot(p-q)\right]\cdot & \underbrace{\left[\Fail(\low,\low-1)+\Fail(\low+1,\low)-\Fail(\low,\low)-\Fail(\low+1,\low-1) \right]}_{K_0}\notag\\
 &&-\eps \cdot &\underbrace{\left[ \Fail(\low+1,\low-1)-\Fail(\low,\low)\right]}_{K_1}<0\notag\\
&&\Leftrightarrow \hspace{0.5cm}-\eps^2\cdot K_0 -\eps\cdot& \underbrace{\left[ (p-q)\cdot K_0+K_1\right]}_L\hspace{1.45cm}<0 \blank.
\end{align}
From \eqref{eq:failure_lemma_2_simple}, which was proven in Lemma~\ref{lem:distance_at_most_2}, it follows that $K_1>0$.
There are three cases.
\begin{compactitem}
 \item[$K_0=0$.] Since we have $K_1>0$, it is easy to see that \eqref{eq:failure_lemma_3_eps} holds for $\eps>0$.
 \item[$K_0>0$.] Regardless whether $L$ is zero, positive, or negative, \eqref{eq:failure_lemma_3_eps} holds for some small $\eps\neq0$.
 \item[$K_0<0$.] The only critical case would be $L=0$, but we will show that it holds $K_1>-K_0$ and therefore $L>0$, 
implying that \eqref{eq:failure_lemma_3_eps} holds for small $\eps>0$.
\end{compactitem}
\smallskip
Inequality $K_1>-K_0$ holds if and only if
\begin{align*}
 \Fail(\low+1,\low)+\Fail(\low,\low-1)>2\cdot \Fail(\low,\low) \blank.
\end{align*}
As before we will simply show the sufficient condition
\begin{align*}
 \fail(\low+1,\low)+\fail(\low,\low-1)>2\cdot \fail(\low,\low) \blank.
\end{align*}
Using \eqref{eq:fail} in combination with the substitutions
$\hf_j=i_j/\cells$  and $\bl=b/\cells$ the condition can be written as
\begin{align*}
(1-\bl)^2\cdot \left[ \bl^{\low-1}-\bl^{2\low-1} \right] >  &\sum_{j=1}^r (1-\bl)^2       \cdot \left[ (\hf_j+\bl)^\low\cdot \bl^{\low-1}-\bl^{2\low-1}\right] \\
                                                    - &\sum_{j=1}^r[1-(\hf_j+\bl)]^2 \cdot \left[ (\hf_j+\bl)^{2\low-1}           -(\hf_j+\bl)^{\low-1}\cdot \bl^\low\right] .
\end{align*}
Note that the subtrahend of the right-hand side is non negative. Hence
it is sufficient to show that
\begin{equation}
\label{eq:failure_lemma_3_x_y}
(1-\bl)^2\cdot \left[ \bl^{\low-1}-\bl^{2\low-1} \right] >    (1-\bl)^2 \cdot\sum_{j=1}^r  (\hf_j+\bl)^\low\cdot \bl^{\low-1} -r\cdot(1-\bl)^2\cdot \bl^{2\low-1} \blank.
\end{equation}
Bounding $\sum_{j=1}^r  (\hf_j+\bl)^\low$ using the binomial theorem gives
\begin{align*}
 \sum_{j=1}^r (\hf_j+\bl)^{\low} &= \sum_{j=1}^r \sum_{i=0}^\low \binom{\low}{i}\cdot \hf_j^i \cdot \bl^{\low-i} = r\cdot \bl^\low +  \sum_{i=1}^\low \binom{\low}{i} \cdot \bl^{\low-i}\cdot \sum_{j=1}^r \hf_j^i \\
&< r\cdot \bl^\low +  \sum_{i=1}^\low \binom{\low}{i}\cdot  \bl^{\low-i}\cdot \Bigg[ \sum_{j=1}^r \hf_j\Bigg]^i 
=(r-1)\cdot \bl^\low +1 \blank,
\end{align*}
where the last step follows from $\sum_{j=1}^r \hf_j=1-\bl$. Substituting $\sum_{j=1}^r (\hf_j+\bl)^{\low}$ with $(r-1)\cdot \bl^\low +1$ 
shows that \eqref{eq:failure_lemma_3_x_y} holds and thus the lemma.\proofEnd

\section{A Conjecture: Essentially Two Different Strategies}
In this section, we give evidence for Conjecture~\ref{con:optimal}, which says that
essentially two types of distributions may be optimal:
one in which all keys are given fixed degrees $\low$ or $\low+1$,
and one in which each node chooses one of $\low$ and $\low+1$ at random, governed by the same
distribution on $\{\low,\low+1\}$. We indicate under what circumstances the one or the other is best.

Assume we are in the situation of Theorem~\ref{theo:main} $(ii)$, i.e., $\low<\Amean<\low+1$ for
some integer constant $\low\geq2$ and it holds $\pmf_x(\low)\in[0,1]$ and $\pmf_x(\low+1)=1-\pmf_x(\low)$, for each $x$ from $S$.
Let $y$ and $z$ be two arbitrary but fixed elements of $S$ with 
\begin{align*}
\pmf_y(\low)=p, \ \pmf_y(\low+1)=1-p && \pmf_z(\low)=q, \ \pmf_z(\low+1)=1-q \blank,
\end{align*}
for $p\in [0,1]$ and $q\in[0,1]$. We would like to know if the matching probability increases
if we change the probability mass functions $\pmf_y$ and $\pmf_z$  to $\pmf'_y$ and $\pmf'_z$, via
\begin{align*}
\pmf'_y(\low)=p+\eps, \ \pmf'_y(\low+1)=1-p-\eps && \pmf'_z(\low)=q-\eps, \ \pmf'_z(\low+1)=1-q+\eps \blank,
\end{align*}
for some $\eps > 0$. 
We note the following.
\begin{compactenum}
\item If $p\geq q$, i.e., $\mean_y\leq \mean_z$, this modification would
move both means towards the boundary of the interval $[\low,\low+1]$. Moving a mean beyond
the boundary cannot increase the matching probability since this would be a contradiction to Lemma~\ref{lem:distance_2}.
\item If $p< q$, i.e., $\mean_y> \mean_z$, this modification would
move both means towards each other.
\end{compactenum}
As in Lemma~\ref{lem:distance_2} it can be shown that the matching probability increases iff
\begin{align*}
   \sum_{\degr_y,\degr_z\in\{\low,\low+1\}} \mgap\Fail(\degr_y,\degr_z) \cdot \rho_y(\degr_y)\cdot\rho_z(\degr_z) 
 &>\mgap \sum_{\degr_y,\degr_z\in\{\low,\low+1\}}\mgap \Fail(\degr_y,\degr_z) \cdot \rho'_y(\degr_y)\cdot\rho'_z(\degr_z) \blank.
\end{align*}
This inequality is equivalent to
\begin{align}
\label{eq:K}
 [-\eps^2 -\eps\cdot(p-q)]\cdot \underbrace{[ \Fail(\low,\low) -2\cdot \Fail(\low,\low+1) + \Fail(\low+1,\low+1) ]}_K < 0 \blank,
\end{align}
utilizing the symmetry $\Fail(\low+1,\low)=\Fail(\low,\low+1)$. Whether inequality \eqref{eq:K} holds or not
depends on $K$, which is independent of $y,z$ and $p,q$. 
There are three cases.
\begin{compactitem}
 \item[$K_0=0$.] The modifications to $\pmf_y$ and $\pmf_z$ do not change the failure probability.
 This case seems unlikely since there would be an infinite number of optimal sequences of probability mass functions;
 hence we will ignore this case for the rest of the discussion.
\item[$K>0$.] Arrange that $p\geq q$ (if necessary interchange $y$ and $z$). Then increasing $p$ and decreasing $q$
 (moving the means away from each other) increases the success probability.
\item[$K<0$.] Arrange that $p<q$ (if $p=q$ do nothing). Again, increasing $p$ and decreasing $q$
(moving the means closer together) increases the success probability.
\end{compactitem}
Using the same method as in Lemmas~\ref{lem:distance_at_most_2}~and~\ref{lem:distance_2} 
it is not possible to show that always $K<0$ or always $K>0$ happens.
To see this, we try to show $K>0$ which is equivalent to proving that
\begin{align}
\label{eq:FAIL_last}
 \Fail(\low,\low)+\Fail(\low+1,\low+1)> 2\cdot \Fail(\low,\low+1) \blank.
\end{align}
As before we only consider the sufficient condition
\begin{align}
\label{eq:fail_last}
\fail(\low,\low)+\fail(\low+1,\low+1)> 2\cdot \fail(\low,\low+1) \blank.
\end{align}
This inequality is equivalent to
\begin{align*}
   \ 2 \cdot \bl^\low     - \bl^{2\low}   + \sum_{j=1}^r [(\hf_j+\bl)^\low    -\bl^\low]^2
 +2 \cdot \bl^{\low+1} - \bl^{2\low+2} + \sum_{j=1}^r [(\hf_j+\bl)^{\low+1}-\bl^{\low+1}]^2 \\
> \hspace{0.5cm} 2 \cdot \bl^\low + 2\cdot \bl^{\low+1} - 2\cdot \bl^{2\low+1} + 2\cdot\sum_{j=1}^r [(\hf_j+\bl)^{\low}-\bl^{\low}]\cdot [(\hf_j+\bl)^{\low+1}-\bl^{\low+1}] \blank,
\end{align*}
where we use the substitutions $\hf_j=i_j/\cells$  and $\bl=b/\cells$. Moving the $\sum_j$-terms to the left and the remaining $\bl$-terms to the right gives
\begin{align*}
 \sum_{j=1}^r \left[ (\hf_j+\bl)^{\low}\cdot (1-\hf_j-\bl) - \bl^\low\cdot(1-\bl)\right]^2 >  \bl^{2\low}\cdot(1-\bl)^2 \blank.
\end{align*}
However, this inequality may hold or may not hold depending on $\hf_j$ and $\bl$. For example, consider
the events 
\begin{compactenum}
 \item $\{\bl=0\}$, then the inequality is true for all $r\geq 2$, and 
 \item $\{r=2$, $\hf_1,\hf_2=1/(2\cdot \low)$, $\bl=1-1/\low\}$, then the inequality is false.
\end{compactenum}
Note that events 1. and 2. have positive probability.

It follows that there exists graphs $\graph_{S-\{y,z\}}$
in which \eqref{eq:fail_last} is true as well as graphs in which \eqref{eq:fail_last} is false.
Hence, it could be possible that there are
nodes $y_1,z_1$ with $K<0$ (their means should be made equal),
and nodes $y_2,z_2$ with $K>0$ (their means should be moved away from each other).
So hypothetically, it could be optimal when 
$S$ is subdivided into 3 disjoint sets $S_{\low}$,$S_{\low+1}$, and $S_{\low,\low+1}$
where each node from $S_{\low}$ has fixed degree $\low$ and each node from $S_{\low+1}$ has fixed degree $\low+1$
and each node from $S_{\low,\low+1}$ has the same mean $\mean\in (\low,\low+1)$, and the degree of each node is concentrated on $\low$ and $\low+1$.
But this would mean if we assume such an ``optimal situation'' and we have three different nodes, say $y_1,y_2$ and $z$, where $y_1,y_2 \in  S_{\low,\low+1}$ and $z\in S_{\low}$,
then it must hold $K>0$ for $\graph_{S-\{y_1,z\}}$ and $K<0$ for $\graph_{S-\{y_1,y_2\}}$ which seems unlikely since $S-\{y_1,z\}$ and $S-\{y_1,y_2\}$ differ in only one node.
(Recall that $K=0$ does not seem plausible, either.) Therefore we conjecture that it is optimal if it holds 
\begin{compactenum}
 \item either    $S=S_{\low} \cup S_{\low+1}$, that is for each $x$ from $S$ the mean $\mean_x$ is fixed to one of the interval borders $\low$ and $\low +1$, and therefore
 a fixed fraction of $\ceil{\Amean}-\Amean$ of the nodes have degree $\low$ (assuming that $\Amean\cdot\keys$ is an integer),
 \item or  $S=S_{\low,\low+1}$, that is it holds $\Amean=\mean_x$ for each $x$ from $S$, and therefore
the number of nodes of degree $\low$ follow a binomial distribution with parameters $n$ and $\ceil{\Amean}-\Amean$.
\end{compactenum}
For the rest of the discussion we only focus on these two degree distributions (fixed and binomial) and 
we try to argue under which conditions case 1 is optimal and when case 2 is optimal.

\begin{figure}
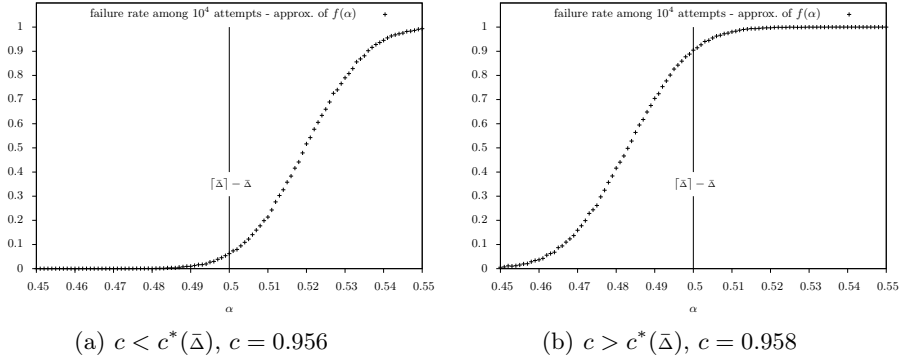

\vspace{-0.8cm}
\hspace{-1.5em}
\subfigure[$c<c^*(\Amean)$, $c=0.956$]{\label{fig:convex} \scalebox{0.5}{\input{\figPath/fixed_load__3-4_c=0956_a=10E4_TEX}}}
\hspace{-1.0em}
\subfigure[$c>c^*(\Amean)$, $c=0.958$]{\label{fig:concave}\scalebox{0.5}{\input{\figPath/fixed_load__3-4_c=0958_a=10E4_TEX}}}
\vspace{-0.5cm}
\caption{\label{fig:convex_concave}Rate of random bipartite graphs with $\randv_x\in\{3,4\}, \Amean=3.5, \cells=10^5,$ 
that have no matching, as a function of $\alpha$ (the fraction of left nodes with degree $3$).
The plots show that the failure function $f(\alpha)$ has probably a transition from convex to concave.
The theoretical threshold $c^*(3.5)$ is approximately $0.957$.}
\vspace{-0.5cm}
\end{figure}

Again our starting point is \eqref{eq:FAIL_last} and $K>0$ respectively. Now fix the degree of all left nodes from $\graph_S$ and
let $\alpha$ be the fraction of nodes from $S$ with degree $\low$ as well as let let $\alpha'$ be the fraction of nodes from $S-\{y,z\}$ with degree $\low$.
Then there are three situations to distinguish according to the degrees of $y$ and $z$.
\begin{compactenum}[$(i)$]
\item $\alpha=\alpha'+2/\keys$, that is $y$ and $z$ have degree $\low$,
\item $\alpha=\alpha'+1/\keys$, that is one node has degree $\low$ the other node has degree $\low+1$,
\item $\alpha=\alpha'$, that is both nodes have degree $\low+1$.
\end{compactenum}
Inequality \eqref{eq:FAIL_last} states that the increase of the failure probability from $(ii)$ to $(i)$ is
larger than the increase of the failure probability from $(iii)$ to $(ii)$ for all $\alpha'$ from $[0,1]$,
that is, the failure probability as a function of $\alpha$ should be convex (while strictly monotonically increasing).
Experimental results as shown in Figure \ref{fig:convex_concave} suggest that this is not the case in general.
In fact there are two different situations for fixed $\Amean$ shown in Figures~\ref{fig:convex}~and~\ref{fig:concave}.
Let $f(\alpha)$ denote the failure probability as a function of $\alpha$.
If $c<c^*(\Amean)$ then
$f$ is convex in a neighborhood of $\ceil{\Amean}-\Amean$. 
Using Jensen's inequality it follows that the failure probability for fixed degree distribution $f(\ceil{\Amean}-\Amean)$ is smaller than 
the failure probability according to the binomial distribution $\sum_{i=0}^\keys f(i/\keys) \cdot \binom{\keys}{i} \cdot (\ceil{\Amean}-\Amean)^i\cdot (1-\ceil{\Amean}+\Amean)^{\keys-i}$,
ignoring the right tail of the binomial distribution that reaches the concave part of $f(\alpha)$, since the tail covers only an exponentially small probability mass.
If $c>c^*(\Amean)$ then $f$ is concave in a neighborhood of $\ceil{\Amean}-\Amean$ and the binomial degree distribution leads to
a smaller failure probability than the fixed degree distribution.

\begin{SCfigure}
\vspace{-0.4cm}
\scalebox{0.5}{\input{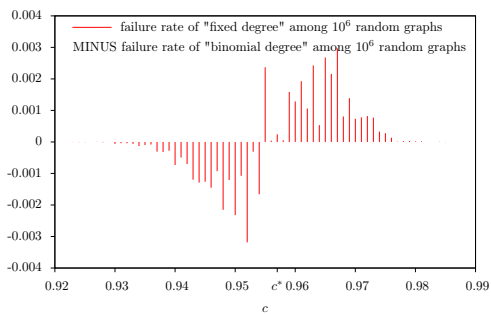}}
\vspace{-0cm}\caption{\label{fig:failure_diff}
Difference of the failure rate of graphs $\graph   \big( \Amean,         ({\pmf}_x)_{x\in S} \big)$
with $\randv_x\in\{3,4\}, \Amean=3.5, \cells=10^3$ and different $({\pmf}_x)_{x\in S}$, as a function of $c=n/m$. 
Minuend is the failure rate using fixed degree, that is $\pmf_x(3)\in\{0,1\}$, for each $x\in S$.
Subtrahend is the failure rate using binomial degree distribution that is $\pmf_x(3)=0.5$, for each $x\in S$.
}  
\end{SCfigure}

In order to back up this observation, an additional experiment was done which directly compares the failure rates for degree distributions
around the threshold. The results are shown in Figure~\ref{fig:failure_diff}.
They confirm the conjecture 
that if $c<c^*(\Amean)$ then the fixed degree distribution is optimal, and if $c>c^*(\Amean)$ then the binomial degree distribution is optimal.
\vspace{-0.7cm}

\vspace{-1cm}
\section{Conclusion}
We found (near) optimal degree distributions for matchings
in bipartite multigraphs where each left node chooses its right neighbors randomly
with repetition according to its assigned degree.
For the case that the number of left nodes is linear in the number
of right nodes we showed that these distributions give matching thresholds
that are the same as the known thresholds for regular/irregular $k$-ary cuckoo hashing;
and in the case of near optimal degree distributions we conjectured the optimal distribution
as a function of the rate of left and right nodes.
\vspace{-0.4cm}
\subsubsection{Acknowledgment.}
The authors would like to thank an anonymous reviewer for pointing out a gap in 
an earlier version of the proof of Lemma~\ref{lem:distance_2}.

\ifnum\arxiv=0
\bibliographystyle{splncs03}
\bibliography{literature.bib}
\else

\fi

\ifnum\arxiv=1
\newpage
\appendix
\section{Degrees Must be Concentrated Around the Mean}
\label{app:lemma_concentration}
In this Section we prove Lemma~\ref{lem:concentration_around_mean}.
Let $(\pmf_x)_{x\in S}$ be given and consider an arbitrary but fixed element $z$ from $S$ with some initial probability mass function $\pmf_z$.
We will show that if there are two degrees of $z$, say $\low$ and $\high$, with non-zero probability and
it holds that $\low < \mean_z < \high$ as well as $\high-\low\geq 2$, then the probability that there is a matching for
the whole key set $S$ cannot be maximal.
More precisely we will show that modifying $\pmf_z$ to $\pmf'_z$ via 
\begin{align*}
\pmf'_z(\low)  &=\pmf_z(\low)-\eps     &\pmf'_z(\high)&=\pmf_z(\high)-\eps \\
\pmf'_z(\low+1)&=\pmf_z(\low+1)+\eps   &\pmf'_z(\high-1)&=\pmf_z(\high-1)+\eps \blank,
\end{align*}
for $\eps \in (0, \min\{\pmf_z(\low),\pmf_z(\high)\}]$, decreases the failure probability,
that is
\begin{align*}
\Pr\left( \evNM_S \mid \pmf_z \right) > \Pr\left( \evNM_S \mid \pmf'_z\right) \blank,
\end{align*}
while it does not change $\mean_z$ and $\Amean$.
For each element $x \in S- \{z\} $ we fix its degree and neighborhood $N_x$.
The resulting graph $\graph_{S-\{z\}}$ can have zero, one or more matchings.
Let $B\subseteq {T}$ be the set of right nodes of $\graph_{S- \{z\}}$ that are matched in every matching
for $S- \{z\}$. Since there can be a matching for $S$ only if there is a matching for $S-\{z\}$ it is sufficient to show that
\begin{align}
\label{eq:failure_lemma_1}
 \Pr\left( \evNM_S \mid \evM_{S-\{z\}}, \pmf_z \right) > \Pr\left( \evNM_S \mid \evM_{S-\{z\}}, \pmf'_z\right) \blank.
\end{align}
Using the law of total probability we get
\begin{align*}
   &\sum_{b=0}^{\keys-1}\Pr\left( \evNM_S \mid \evM_{S-\{z\}}, \pmf_z,  \abs{B}=b \right)\cdot \Pr\left(\abs{B}=b \mid \evM_{S-\{z\}}, \pmf_z \right)\\
 > &\sum_{b=0}^{\keys-1}\Pr\left( \evNM_S \mid \evM_{S-\{z\}}, \pmf'_z, \abs{B}=b \right)\cdot \Pr\left(\abs{B}=b \mid \evM_{S-\{z\}}, \pmf'_z \right)\blank.
\end{align*}
In order that $\graph_S$ has a matching there must be at least one node in the neighborhood $N_z$ of $z$ that is not an element of $B$. Therefore
we have to show
\begin{align*}
  &\sum_{b=0}^{\keys-1}\left[\sum_{d=1}^\cells \pmf_z(d)\cdot \bfrac{b}{\cells}^d \right] \cdot \Pr\left(\abs{B}=b \mid \evM_{S-\{z\}}, \pmf_z \right)\\
> &\sum_{b=0}^{\keys-1}\left[\sum_{d=1}^\cells \pmf'_z(d)\cdot \bfrac{b}{\cells}^d\right]\cdot \Pr\left(\abs{B}=b \mid \evM_{S-\{z\}}, \pmf'_z \right) \blank.
\end{align*}
Note that $B$ is independent of $z$ and $\pmf_z$, respectively, and if $b=0$ the modification from $\pmf_z$ to $\pmf'_z$ does not affect the failure probability.
Hence we consider only the cases where $b>0$ and it remains to show
\begin{align*}
 & &\sum_{d=1}^\cells \pmf_z(d)\cdot \bfrac{b}{\cells}^d 
&> \sum_{d=1}^\cells \pmf'_z(d)\cdot \bfrac{b}{\cells}^d\\
\Leftrightarrow & &
\eps \cdot \bfrac{b}{\cells}^\low+\eps \cdot \bfrac{b}{\cells}^\high &> \eps \cdot \bfrac{b}{\cells}^{\low+1}+\eps \cdot \bfrac{b}{\cells}^{\high-1 }\\
\Leftrightarrow& &
\bfrac{b}{\cells}^{\low}\cdot\left(1-\frac{b}{\cells}\right)  &> \bfrac{b}{\cells}^{\high-1}\cdot\left(1-\frac{b}{\cells}\right) \blank, 
\end{align*}
which is true since $0<b/\cells<1$, $\high-\low \geq 2$. Since the event $\{b>0\}$ has positive probability,
inequality \eqref{eq:failure_lemma_1} holds. This finishes the proof of Lemma~\ref{lem:concentration_around_mean}.~\proofEnd

\section{Asymptotic Behavior and Thresholds}
\label{app:proposition}
In this section we give the proof of Proposition~\ref{prop:thresholds}.
Let $n=c\cdot m$ for $c>0$ and let $({\pmf}_x)_{x\in S}$ be a near optimal sequence of degree distributions with
$\Amean=\alpha\cdot\floor{\Amean} +(1-\alpha)\cdot \ceil{\Amean}>2$ for  $\alpha\in(0,1]$.
Consider the random graph $\graph\big( \Amean, ({\pmf}_x)_{x\in S} \big)$
where each left node has $\randv_x\in \{\floor{\Amean},\ceil{\Amean}\}$ random neighbors
(not necessarily distinct) and $\randv_x$ is distributed according to $\pmf_x$
where it holds $\alpha=1/{\keys}\cdot \sum_{x\in S}\pmf_x(\floor{\Amean})$.

Let $\low=\floor{\Amean}$. We consider a new random bipartite graph $\graphWR(\low,\alpha)$ 
with $\keys$ left nodes and $\cells$ right nodes where 
a constant fraction of $\alpha$ left nodes has 
degree $\low$, a fraction of $1-\alpha$ left nodes has degree $\low+1$,
and the neighbors of each left node are chosen uniformly at random without replacement.
In summary, for $\graph$ the degrees of the left nodes are randomly chosen and
duplicate neighbors are allowed; for $\graphWR$ the degrees of the left nodes are fixed and
the neighbors are pairwise distinct.

Now, for each $x$ from $S$ let $Y^\low_x$ be a binary random variable
with $Y^\low_x=1$, if the neighborhood set $N_x$ of $x$ has size $\low$ and $Y^\low_x=0$, if $N_x$ has size strictly smaller than $\low$.
Furthermore let $Y^\low=\sum_{x\in S}Y^\low_x$. Then
\begin{align}
\label{eq:expected_degree}
 \E(Y^\low_x)
&=\pmf_x(\low)\cdot \frac{\binom{\cells}{\low}\cdot \low!}{ \cells^\low }
 +( 1-\pmf_x(\low))\cdot \frac{\binom{\cells}{\low}\cdot \low! \cdot \binom{\low+1}{2}}{\cells^{\low+1}} \text{, and} \\
 \E(Y^{\low+1}_x)&=( 1-\pmf_x(\low))\cdot \frac{\binom{\cells}{\low+1}\cdot (\low+1)!}{ \cells^{\low+1} }\notag \blank.
\end{align}
Consider the events
\begin{compactenum}
 \item $\evA=\{ \keys \cdot \alpha     -n^\delta \leq  Y^\low \leq \keys\cdot \alpha +n^\delta \}$ and
 \item $\evB=\{ \keys \cdot (1-\alpha) -n^\delta \leq  Y^{\low+1} \leq \keys\cdot(1- \alpha) +n^\delta \}$, 
 \end{compactenum}
stating that the number of left nodes with neighborhood size  $\low$ and $\low +1$ is near $\keys \cdot \alpha$ and $\keys \cdot (1-\alpha)$, respectively.

We want to bound the probability of $\Pr(\evA \cup \evB)$ using the complementary event $\bar{\evA} \cap \bar{\evB}$,
via $\Pr( \bar{\evA} \cap \bar{\evB})\leq \Pr(\bar{\evA})+\Pr(\bar{\evB})$.

First consider the event $\bar{\evA}$. Let $Y_x=Y^\low_x$ and let $Y=Y^\low$ as well as let $p_x=\Pr(Y_x=1)$.
According to \eqref{eq:expected_degree} it holds that 
\begin{align*}
p_x=\E(Y_x)=\pmf_x(\low) \cdot \big( 1-\Theta(1/\cells) \big) + \Theta(1/\cells) \blank,
\end{align*}
since  $1-{\low^2}/{\cells}<{\binom{\cells}{\low}\cdot \low!}/{ \cells^\low }< 1-{1}/{\cells}$,
where the lower bound follows from Bernoulli's inequality.

For each $x \in S$ let $Z_x=Y_x-p_x$. Now fix an arbitrary order of the left nodes, i.e., let $S=\{x_1,x_2,\ldots,x_n\}$.
It holds that $X_0,X_1,\ldots,X_n$ with $X_0=0$ and $X_i=X_{i-1}+Z_{x_i}$
is a martingale with bounded differences, since 
\begin{align*}
\E(X_{i+1} \mid X_0,\ldots,X_i)= \E(X_{i}+Z_{x_{i+1}} \mid X_0,\ldots,X_i)=X_i 
\end{align*}
and  $\abs{X_{i+1}-X_i}\leq 1$. Applying a standard Azuma–Hoeffding inequality~\cite[Theorem 5.1]{DP_concenctraion_2009} we get
\begin{align*}
 \Pr\left( \abs{X_n-X_0}\geq n^\gamma \right)=\Pr\left( \abs{ Y - \E(Y)}\geq n^\gamma \right)\leq 2\cdot e^{{-2\cdot n^{2\cdot \gamma}}/{n}} \blank.
\end{align*}
That is for $\gamma>1/2$ the probability that number of left nodes that have a neighborhood set of size $\low$ 
differ more than $n^\gamma$ from its expected value is exponentially small in $\keys$. 
By linearity of expectation, it holds 
\begin{align*}
 \E(Y)=\sum_{x \in S}p_x= \big(1-\Theta(1/m) \big) \cdot \sum_{x\in S}\pmf_x(\low) + \Theta(1) \blank,
\end{align*}
and since $\alpha=1/\keys\cdot \sum_{x\in S} \pmf_x(\low)$ it follows that $\E(Y)=\keys \cdot \alpha \pm \Theta(1)$.
Thus, one can conclude that if $1>\delta>\gamma>1/2$ then the probability of event $\bar{\evA}$ is exponentially small in $\keys$.
Essentially the same proof shows an exponentially small bound for~$\bar{\evB}$.

Hence the probability of the event $\evM[ \graph   \big( \Amean,         ({\pmf}_x)_{x\in S} \big)]$ that $\graph$
has a left-perfect matching can be bounded via
\begin{equation*}
\begin{split}
\Pr\big( \evM[ &\graph   \big( \Amean,         ({\pmf}_x)_{x\in S} \big)] \big) \\
 &=\Pr\left( \evM[ \graph   \big( \Amean,         ({\pmf}_x)_{x\in S} \big)] \mid \evA \cup \evB \right) \cdot \Big(1-O\big(e^{-\keys^{2\delta-1}}\big)\Big)+O\big(e^{-\keys^{2\delta-1}}\big) \blank.
\end{split}
\end{equation*}
Now consider the graph $\graphWR (\low, \alpha )$. 
From \cite[Theorem 3]{DGMMPR_tight_2010} it follows with similar concentration bounds as above that
there is a constant $c^*(\low,\alpha)$ such that for $n\to \infty$ we have,
if $c={\keys}/{\cells}<c^*(\low,\alpha)$ then the probability
that $\graphWR (\low, \alpha)$ has a matching goes to $1$ and
if $c>c^*(\low,\alpha)$ then the probability that $\graphWR (\low, \alpha)$ has a matching  goes to $0$.
The point of transition from success to failure is exactly the point where the $2$-core of the corresponding
hypergraph, which is the largest induced sub-hypergraph that has minimum degree at least $2$,
gets edge density larger than $1$; see e.g.~\cite{FM_maximum_2009,FP_orientability_2010} for the case of
hyperedges of only one size and \cite{DGMMPR_tight_2009_full,DGMMPR_tight_2010} for the generalization to
hyperedges of different sizes. If the $2$-core is not empty then its number of edges
is linear in $\keys$ and its number of nodes is linear in $\cells$. Now assume that the event $\evA\cup \evB$ takes place.
Let $\graph'$ be the induced subgraph of $\graph$ that covers each left node
and its neighborhood if the left node has either $\low$ or $\low+1$ pairwise distinct
neighbors.
The $2$-core of the hypergraph regarding $\graph'$ has asymptotically the same density as
the $2$-core of the hypergraph regarding $\graphWR$. 
But since the $2$-core has linear size or is empty it follows that the $2$-core of $\graph$ has asymptotically the same density too.
Hence the proposition follows.\proofEnd
\fi

\def\addendum{0}
\ifnum\addendum=1
\section{Addendum to Theorem~\ref{theo:main}~$(ii)$}
Consider the situation $n=m^\delta$, for fixed $\delta\in(0,1/2)$.
Let $y$ and $z$ be two arbitrary but fixed elements of $S$ with $\mean_y\leq \mean_z$ and
\begin{align*}
\pmf_y(\low)&=p  & \pmf_y(\low+1)=&1-p\\
\pmf_z(\low)&=q  & \pmf_z(\low+1)=&1-q \blank,
\end{align*}
for $p,q \in [0,1]$ and $p\geq q$. We will show that the matching probability increases
if we change the probability mass functions $\pmf_y$ and $\pmf_z$  to $\pmf'_y$ and $\pmf'_z$ where
\begin{align*}
\pmf'_y(\low)=&p+\eps & \pmf'_y(\low+1)=&1-p-\eps\\
\pmf'_z(\low)=&q-\eps & \pmf'_z(\low+1)=&1-q+\eps \blank,
\end{align*}
for some $\eps > 0$. 
Note that since $\mean_y\leq \mean_z$ this modification moves both means towards the boundary of the interval $[\low,\low+1]$. Moving a mean beyond
the boundary cannot increase the matching probability since this would be a contradiction to Lemma~\ref{lem:distance_2}.
The matching probability increases if and only if
\begin{align*}
   \sum_{\degr_y,\degr_z\in\{\low,\low+1\}} \mgap \Fail(\degr_y,\degr_z) \cdot \rho_y(\degr_y)\cdot\rho_z(\degr_z) 
 &>\mgap\sum_{\degr_y,\degr_z\in\{\low,\low+1\}} \mgap\Fail(\degr_y,\degr_z) \cdot \rho'_y(\degr_y)\cdot\rho'_z(\degr_z) \blank.
\end{align*}
This inequality is equivalent to
\begin{align}
 [-\eps^2 -\eps\cdot(p-q)]\cdot \underbrace{[ \Fail(\low,\low) -2\cdot \Fail(\low,\low+1) + \Fail(\low+1,\low+1) ]}_K < 0 \blank,
\end{align}
utilizing the symmetry $\Fail(\low+1,\low)=\Fail(\low,\low+1)$. Since $p\leq q$ the inequality holds 
if and only if $K>0$, where $K$ is independent of $y,z$ and $p,q$, respectively.
To show $K>0$ that is 
\begin{align}
 \Fail(\low,\low)+\Fail(\low+1,\low+1)> 2\cdot \Fail(\low,\low+1) \blank,
\end{align}
we simply prove the the sufficient condition
\begin{align}
\fail(\low,\low)+\fail(\low+1,\low+1)> 2\cdot \fail(\low,\low+1) \blank.
\end{align}
This inequality is equivalent to
\begin{align*}
 \sum_{j=1}^r \left[ (\hf_j+\bl)^{\low}\cdot (1-\hf_j-y) - \bl^\low\cdot(1-\bl)\right]^2 >  \bl^{2\low}\cdot(1-\bl)^2 \blank,
\end{align*}
where $\hf_j=i_j/\cells$ and $\bl=b/\cells$.
As before we are only interested in the cases where $b\geq 1$.
Division by $\bl^{2\low}\cdot(1-\bl)^2$ gives
\begin{align*}
 \sum_{j=1}^r \left[ \left(1+\frac{\hf_j}{\bl}\right)^{\low}\cdot \left(1-\frac{\hf_j}{1-\bl}\right) - 1\right]^2 >  1 \blank.
\end{align*}
Now observe that there are at least $\cells-(\low+1)\cdot \keys=\cells-(\low+1)\cdot \cells^\delta$ free nodes,
that is summands with $\hf_j=1/\cells$.
Hence the inequality holds if
\begin{align*}
  \left[\cells-(\low+1)\cdot \cells^\delta \right]\cdot \Bigg[ \underbrace{\left(1+\frac{1}{b}\right)^{\low}\cdot \left(1-\frac{1}{m-b}\right)}_{f(\low)} - 1\Bigg]^2 >  1 \blank.
\end{align*}
The function $f(\low)$ grows for growing $\low$ and since $1\leq b\leq n=m^\delta$ we have that $f(1)>1$.
Therefore it is sufficient to show the last inequality for $\low=1$, that is
\begin{align*}
  \left[\cells-(\low+1)\cdot \cells^\delta \right]\cdot \underbrace{\Bigg[ \frac{\cells-2\cdot b-1}{b\cdot(m-b)}\Bigg]^2}_{g(b)} >  1 \blank.
\end{align*}
Since $g(b)>1/(4b^2)$ and $b^2<m$ the inequality holds.
\fi
\end{document}